\documentclass{aastex631}

\begin{document}

\title{BSN: The First Light Curve Analysis of the Total Eclipse Binary System EL Tuc}

\author{Elham Sarvari}
\affiliation{Independent researcher, Tehran, Iran; fh.elhamsarvari94@gmail.com}

\author{Eduardo Fernández Lajús}
\affiliation{Instituto de Astrofísica de La Plata (CCT La Plata-CONICET-UNLP), La Plata, Argentina}

\author{Atila Poro}
\affiliation{BSN Project, \& Astronomy Department of the Raderon AI Lab., BC., Burnaby, Canada}

\begin{abstract}
We conducted the first light curve study of the binary star EL Tuc within the Binary Systems of South and North (BSN) Project framework. The photometric observations were made using standard multiband $BVR_cI_c$ filters at an observatory in Argentina. We presented a new ephemeris for EL Tuc and a linear fit to the O-C diagram, utilizing our extracted times of minima and additional literature. We employed the PHysics Of Eclipsing BinariEs (PHOEBE) Python code and the Markov chain Monte Carlo (MCMC) approach for the system's light curve analysis. The target system's light curve solution required a cold starspot on the hotter component. We conclude that EL Tuc is a total contact binary system with a low mass ratio of $q=0.172\pm0.002$, an orbital inclination of $i=83.74 \pm0.40$ degree, and a fillout factor of $f=53.7\pm1.6\%$. We used the $P-a$ relationship and the Gaia Data Release 3 (DR3) parallax method to determine the absolute parameters of EL Tuc to compare the precision of our results. This system was classified as W-type based on the mass and effective temperature of the companion stars. The positions of the systems were depicted on the $M-L$, $M-R$, $T-M$, and $q-L_{ratio}$ diagrams. The relationship between the spectroscopic and photometric mass ratios of binaries was discussed.
\end{abstract}


%%%%%%%%%%%%%%%%% BODY OF PAPER %%%%%%%%%%%%%%%%%%
\section{Introduction}
Eclipsing binaries play a crucial role in astrophysics by providing valuable insights into star formation, stellar structure, and stars' physical characteristics and evolution (\citealt{stassun2008eclipsing}, \citealt{southworth2012eclipsing}, \citealt{stassun2014empirical}). According to the categories of eclipsing binaries provided by \cite{1959cbs..book.....K} based on the Roche geometry, they were classified as detached, semi-detached, and contact systems. The W Ursae Majoris (W UMa) systems typically consist of F, G, or K spectral-type stars that have filled their Roche lobes (\citealt{van1982evolutionary}). When a star fills its Roche lobe, its companion leads to mass and energy exchange (\citealt{paczynski1971evolutionary}, \citealt{yakut2005evolution}, \citealt{paczynski2006eclipsing}, \citealt{zhang2016first}). These systems exhibit continuous variations in brightness and nearly identical depths in both eclipse minima, indicating that the two components have almost the same temperatures (\citealt{lucy1968structure}, \citealt{lucy1976w}). W UMa contact systems typically have orbital periods of $P_{orb}\leq1$ days, with most of them in the range of approximately 0.2 to 0.6 days (\citealt{dryomova2006variation}, \citealt{latkovic2021statistics}). Although inspection of All-Sky Survey data reveals that contact binary stars are common \cite{rucinski20027}, our comprehension of their origin, structure, and evolution remains incomplete. Consequently, the challenge lies in developing a satisfactory theory and conducting further studies to explain these kinds of binary systems.

The W UMa-type systems are the most common low-mass systems. Among them, systems with a low mass ratio are also interesting to investigate. Such systems suggest the possibility of star mergers leading to the formation of blue stragglers, red novae, and fast-rotating stars (\citealt{2024SerAJ.208....1A}). Also, systems with extremely low mass ratios are still being discovered and studied.

Contact binary systems are further classified into two subtypes, A and W-types, based on the companion stars' mass and effective temperature (\citealt{binnendijk1970orbital}). A-type systems, exhibiting slightly stronger contact in their envelopes than W-type systems, predominantly have periods exceeding 0.41 days, as demonstrated by \cite{qian2003overcontact}. In A-type systems, the more massive component has a higher effective temperature. 
\\
\\
EL Tuc, a binary system located in the southern hemisphere and Tucana constellation, has an apparent magnitude range of $V=14.48–14.94$ (VSX\footnote{The International Variable Star Index, \url{https://www.aavso.org/vsx/}}) and an orbital period of 0.33720 days (\citealt{paschke2007list}). The system's coordinate from Gaia DR3\footnote{\url{https://gea.esac.esa.int/archive/}} is RA.: $0.26796499^{\circ}$ and Dec.: $-66.96201565^{\circ}$. According to existing catalogs (e.g., APASS9 \citealt{vallenari2023gaia}, and ASAS \citealt{pojmanski2002all}), EL Tuc is classified as a contact binary system. So far, EL Tuc (GSC 08846-00581) has not been the main subject of any literature for light curve solutions.
\\
\\
This investigation aims to present the results of our first light curve analysis of the EL Tuc binary system and identify its characteristics. The study is organized as follows: Section 2 provides details on the multi-color CCD light curves performed at an observatory in Argentina and a data reduction process; Section 3 outlines the process of extracting times of minima and calculating a new ephemeris of EL Tuc. The light curve analysis for our system is discussed in Section 4. Section 5 presents the estimation of the system’s absolute parameters, and finally, a discussion and conclusion are given in Section 6. This study continues the BSN\footnote{\url{https://bsnp.info/}} project's observations and analyses of contact binary systems.

%%%%%%%%%%%%%%%%%%%%%%%%%%%%%%%%%%%%%%%%%%%%%%%%%%
\vspace{1cm}
\section{Observation and Data Reduction}
Photometric observation of EL Tuc was executed in September 2023, capturing 652 images in one night. This observation was made using the 2.15-meter Jorge Sahade (JS) telescope, located at the Complejo Astronomico El Leoncito (CASLEO) Observatory in Argentina ($69^{\circ}$ $18^{\prime}$ W, $31^{\circ}$ $48^{\prime}$ S, $2552$ meter above sea level).
A Roper scientific cryogenic CCD, VersArray 2048B, and $BVR_cI_c$ standard filters were employed. Each frame was $5\times5$ binned with an exposure time of 50 seconds for the $B$ filter, 20 seconds for the $V$ filter, 20 seconds for the $R_c$ filter, and 20 seconds for the $I_c$ filter.

Gaia DR3 4707675346333770496 (RA.: $00^h$ $01^m$ $01.30^s$, Dec.: $-66^\circ$ $57'$ $21.33"$(J2000)) and Gaia DR2 4707675861729845248 (RA.: $00^h$ $01^m$ $28.27^s$, Dec.: $-66^\circ$ $58'$ $37.59"$(J2000)) were selected as comparison and check stars, respectively.

The APPHOT photometry package of the Image Reduction and Analysis Facility2 (IRAF) was used for CCD reduction and aperture photometry (\citealt{tody1986iraf}). The basic data reduction was carried out, which included bias, and flat fielding for each CCD image. For all of the observation data, we used Python code based on the Astropy package (\citealt{2013AA...558A..33A}) to consider the airmass with a formula from the \cite{1962aste.book.....H} study.
Then, the flux of the data was normalized by employing the AstroImageJ program (\citealt{2017AJ....153...77C}).

It should be mentioned that this system was observed only in sector 1 with an 1800-second exposure time of the Transiting Exoplanet Survey Satellite (TESS), and the data were unsuitable for this investigation. Therefore, the ground-based data are currently useful in studying this system.

%%%%%%%%%%%%%%%%%%%%%%%%%%%%%%%%%%%%%%%%%%%%%%%%%%
\vspace{1cm}
\section{New Ephemeris}
We extracted four primary and four secondary times of minima from our observations and gathered thirteen other minima from the literature. These times of minima are listed in Table \ref{tab1}. Barycentric Julian Date in Barycentric Dynamical Time $(BJD_{TDB})$ was used to express all times of minima and perform the calculations process. We considered the following light elements as a reference ephemeris from the \cite{paschke2007list} study for computing the epoch and O-C values (Equation \ref{eq1}),

\begin{equation}\label{eq1}
Min.I(BJD_{TDB})=(2454275.63275\pm0.00200)+(0.33720)\times E
\end{equation}

where $E$ is the number of cycles. 

The \cite{juryvsek2017brno} study reported $2456966.71592\pm0.00030$ as a primary minimum. However, using our ephemeris reference (Equation \ref{eq1}), we recognized this minimum as a secondary type. The ephemeris reference we used is in agreement with determining the types of all collected times of minima, as supported by their reference studies. This implies that the minimum time reported by \cite{juryvsek2017brno} might not have been accurate enough, or they may have been calculated using a different ephemeris reference. So, although we have shown this minimum time in the O-C diagram, and due to its being far from other points, it was not used to calculate the new ephemeris. Also, a secondary minimum time from \cite{hovnkova2013brno} was also ignored due to the point outlier in the O-C diagram.

\begin{figure}
\begin{center}
\includegraphics[width=0.7\textwidth]{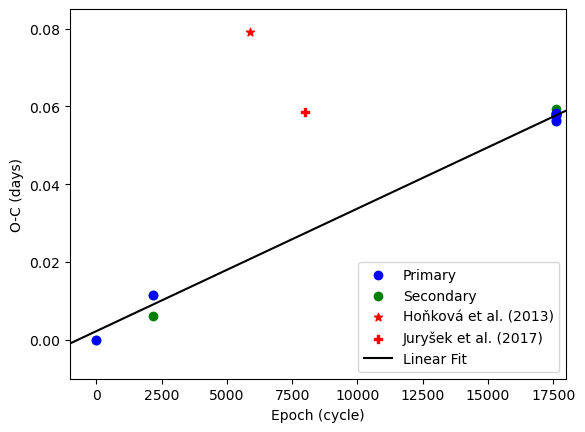}
\caption{The O-C diagram of the EL Tuc binary system. The red outliers are not used in the ephemeris calculation and are explained in the new ephemeris section.}
\label{Fig1}
\end{center}
\end{figure}

Table \ref{tab1} is structured such that the first column represents times of minima, their uncertainties, epochs, and O-C values in the subsequent columns, with the final column containing the references. The O-C diagram, derived from the primary times of minima, was plotted as shown in Figure \ref{Fig1}. The O-C diagram displays a linear least squares fit to the points. So, we calculated a new ephemeris for the EL Tuc system (Equation \ref{eq2}).

\begin{equation}\label{eq2}
Min.I(BJD_{TDB})=(2454275.63499\pm0.00140)+(0.33720315\pm0.00000010)\times E.
\end{equation}

\begin{table*}
\caption{Extracted and collected CCD-observed times of minima.}
\centering
\begin{center}
\footnotesize
\begin{tabular}{c c c c c}
\hline
\hline
Min.($BJD_{TDB}$) & Error & Epoch & O-C(day) & Reference\\
\hline
2454275.63275	&	0.00200	&	0	&	0	&	\cite{paschke2007list}	\\
2455004.49675	&	0.00700	&	2161.5	&	0.0062	&	\cite{paschke2009list}	\\
2455004.67075	&	0.01000	&	2162	&	0.0116	&	\cite{paschke2009list}	\\
2456263.67441	&	0.00070	&	5895.5	&	0.0791	&	\cite{hovnkova2013brno}	\\
2456966.71592	&	0.00030	&	7980.5	&	0.0586	&	\cite{juryvsek2017brno}	\\
2460209.56663	&	0.00036	&	17597.5	&	0.0569	&	This study	\\
2460209.56715	&	0.00030	&	17597.5	&	0.0574	&	This study	\\
2460209.56763	&	0.00035	&	17597.5	&	0.0579	&	This study	\\
2460209.56911	&	0.00049	&	17597.5	&	0.0594	&	This study	\\
2460209.73463	&	0.00034	&	17598	&	0.0563	&	This study	\\
2460209.73623	&	0.00020	&	17598	&	0.0579	&	This study	\\
2460209.73633	&	0.00024	&	17598	&	0.0580	&	This study	\\
2460209.73656	&	0.00022	&	17598	&	0.0582	&	This study	\\
\hline
\hline
\end{tabular}
\end{center}
\label{tab1}
\end{table*}

%%%%%%%%%%%%%%%%%%%%%%%%%%%%%%%%%%%%%%%%%%%%%%%%%%
\vspace{1cm}
\section{Light Curve Solution}
We conducted the first light curve analysis for the EL Tuc binary system. The PHOEBE Python code version 2.4.9 (\citealt{prvsa2016physics}, \citealt{conroy2020physics}), and the MCMC approach were employed. According to the shape of the light curve and target system type in catalogs, we used contact mode for light curve analysis. The bolometric albedo and gravity-darkening coefficients were assumed to be $A_1=A_2=0.5$ (\citealt{rucinski1969proximity}) and $g_1=g_2=0.32$ (\citealt{lucy1967gravity}) respectively. We used the \cite{castelli2004new} study to describe the stellar atmosphere and adopted the limb darkening coefficients as free parameters in the PHOEBE code. The hotter component's initial effective temperature was set from Gaia DR3\footnote{\url{https://www.cosmos.esa.int/web/gaia/data-release-3}}.

\begin{figure}
\begin{center}
\includegraphics[width=0.68\textwidth]{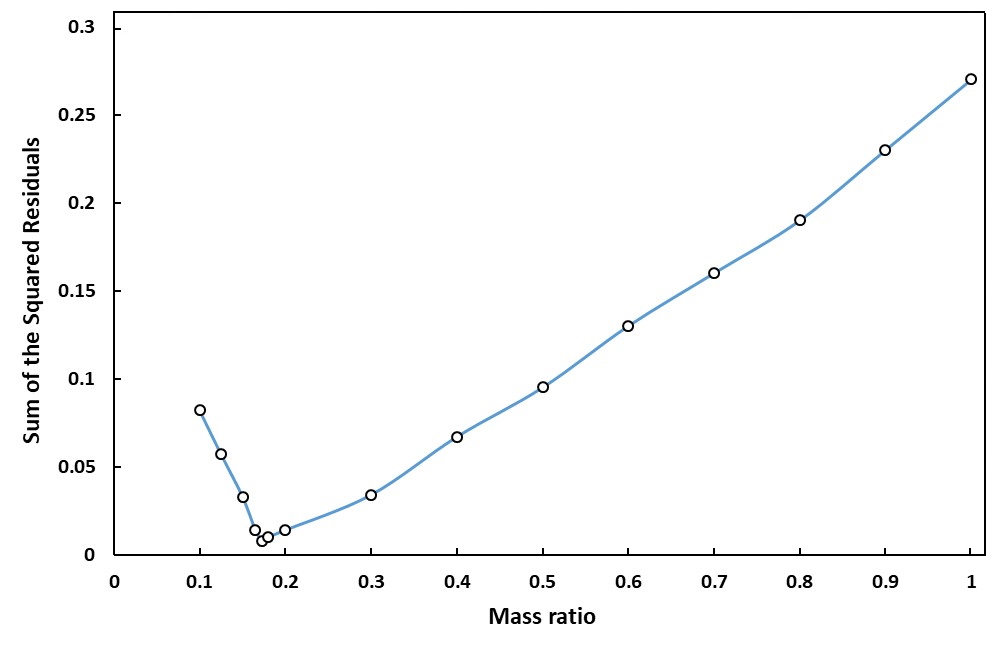}
\caption{Sum of the squared residuals as a function of the mass ratio.}
\label{Fig2}
\end{center}
\end{figure}

The $q$-search method was used to estimate the initial system's mass ratio ($q=M_2/M_1$) since the photometric data were available. The process of determining the initial mass ratio was first carried out in the range of 0.1 to 10 and then between 0.1 and 0.9 with shorter steps (Figure \ref{Fig2}). The determined mass ratio showed that EL Tuc is one of the systems with a low mass ratio (\citealt{2022AJ....164..202L}).

The O'Connell effect (\citealt{o1951so}), which refers to the unequal light level between the primary and secondary maxima, is well-known in numerous eclipsing binaries. This effect is typically attributed to surface inhomogeneities on one or both of the components and is modeled by positioning hot or cold starspot(s) on these components. In this study, the best solution was found with a cold starspot on the hotter component, and its parameters are shown in Table \ref{tab2}.

Ultimately, we utilized the MCMC approach based on the emcee package (\citealt{2013PASP..125..306F}) to enhance precision in our modeling, thereby improving the results of the light curve solutions and obtaining the final results. In the MCMC process, we used 24 walkers, each undergoing 1000 iterations. Therefore, six main parameters, including $i$, $q$, $f$, $T_{1,2}$, and $l_1$, along with four starspot parameters (coordinates, radius, and $T_{spot}/T_{star}$), were considered for the MCMC modeling process. The MCMC process is known to be used to obtain reasonable upper and lower limit uncertainties for each parameter.

\begin{figure*}
\begin{center}
\includegraphics[width=0.8\textwidth]{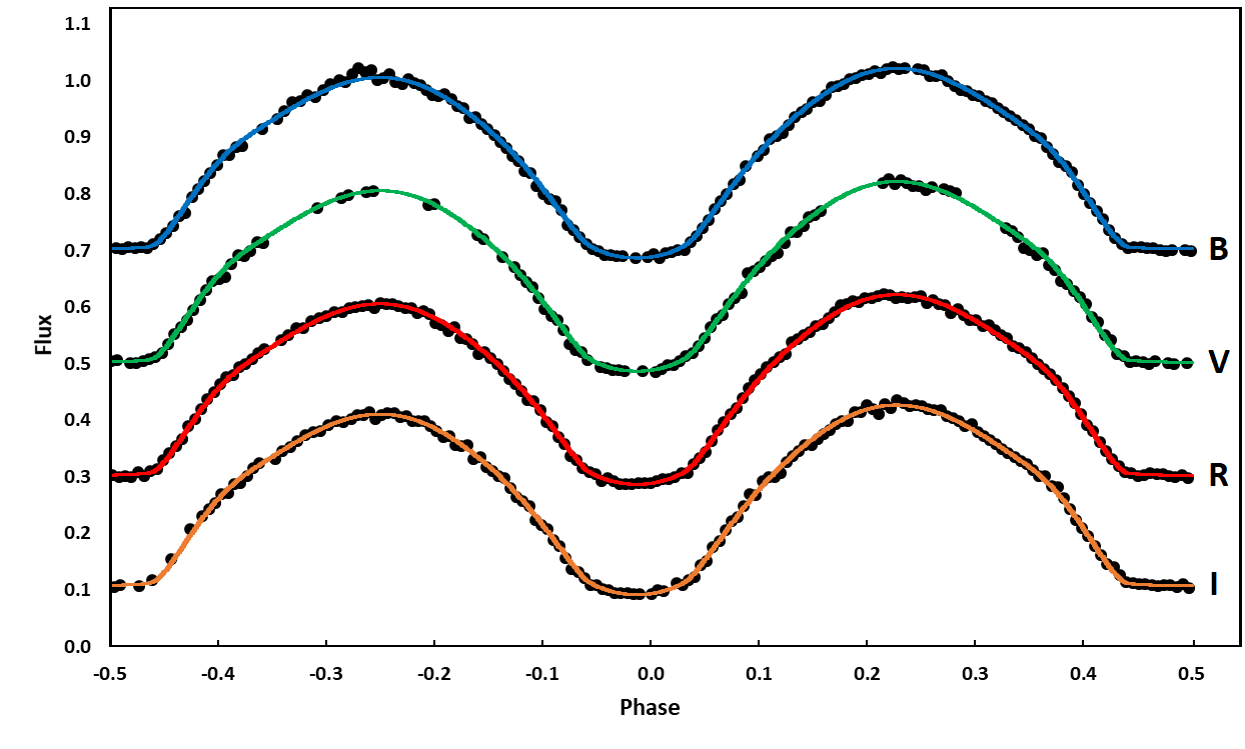}
\caption {The observed light curves of EL Tuc, represented by points, and the modeled solutions, represented by lines, are shown in the $BVR_cI_c$ filters. The curves are arranged from top to bottom based on the orbital phase and have been shifted arbitrarily in the relative flux.}
\label{Fig3}
\end{center}
\end{figure*}

\begin{figure*}
\begin{center}
\includegraphics[width=0.98\textwidth]{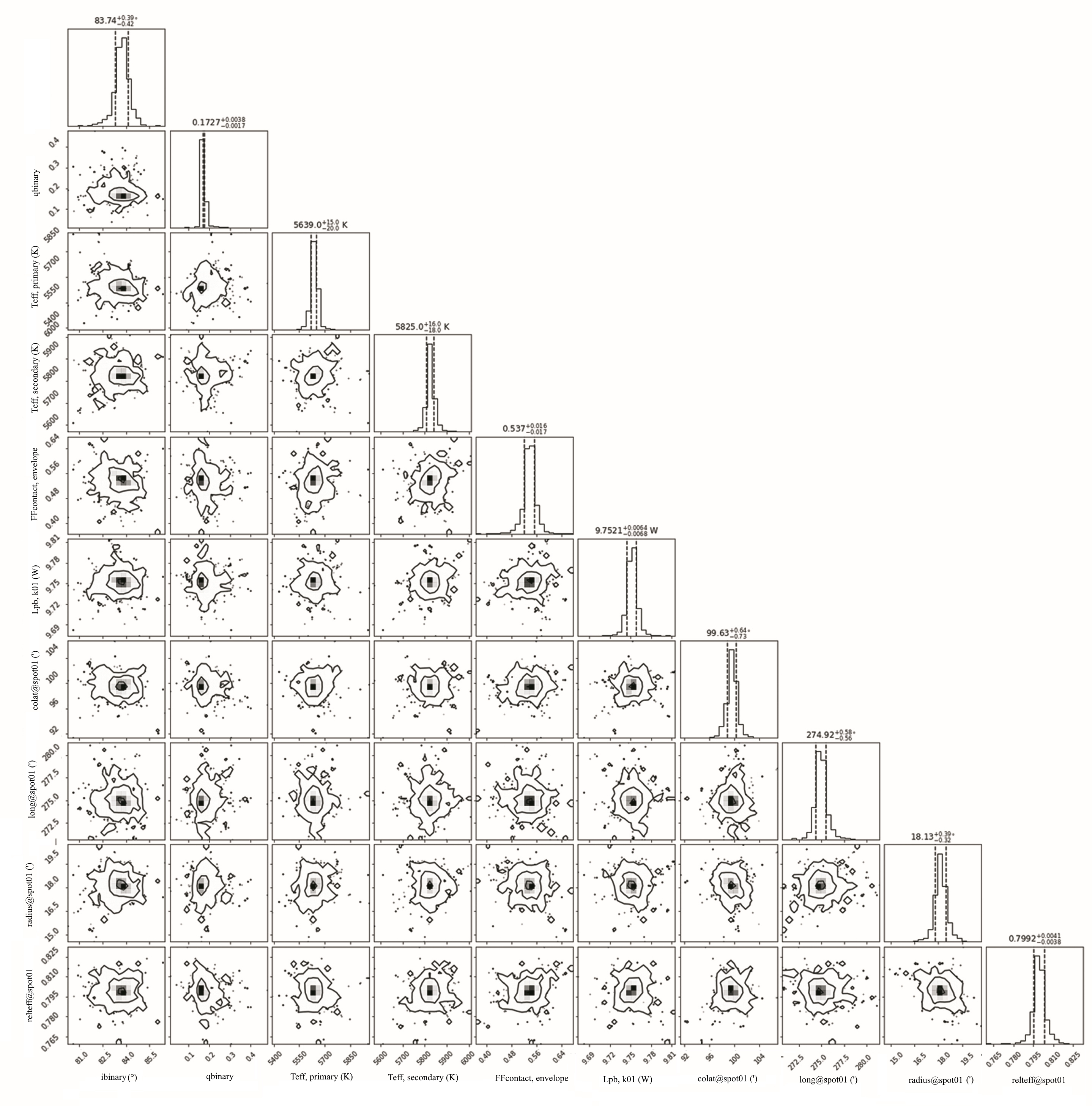}
\caption {The corner plots of the EL Tuc system were determined by MCMC modeling.}
\label{Fig4}
\end{center}
\end{figure*}

\begin{figure*}
\begin{center}
\includegraphics[width=0.72\textwidth]{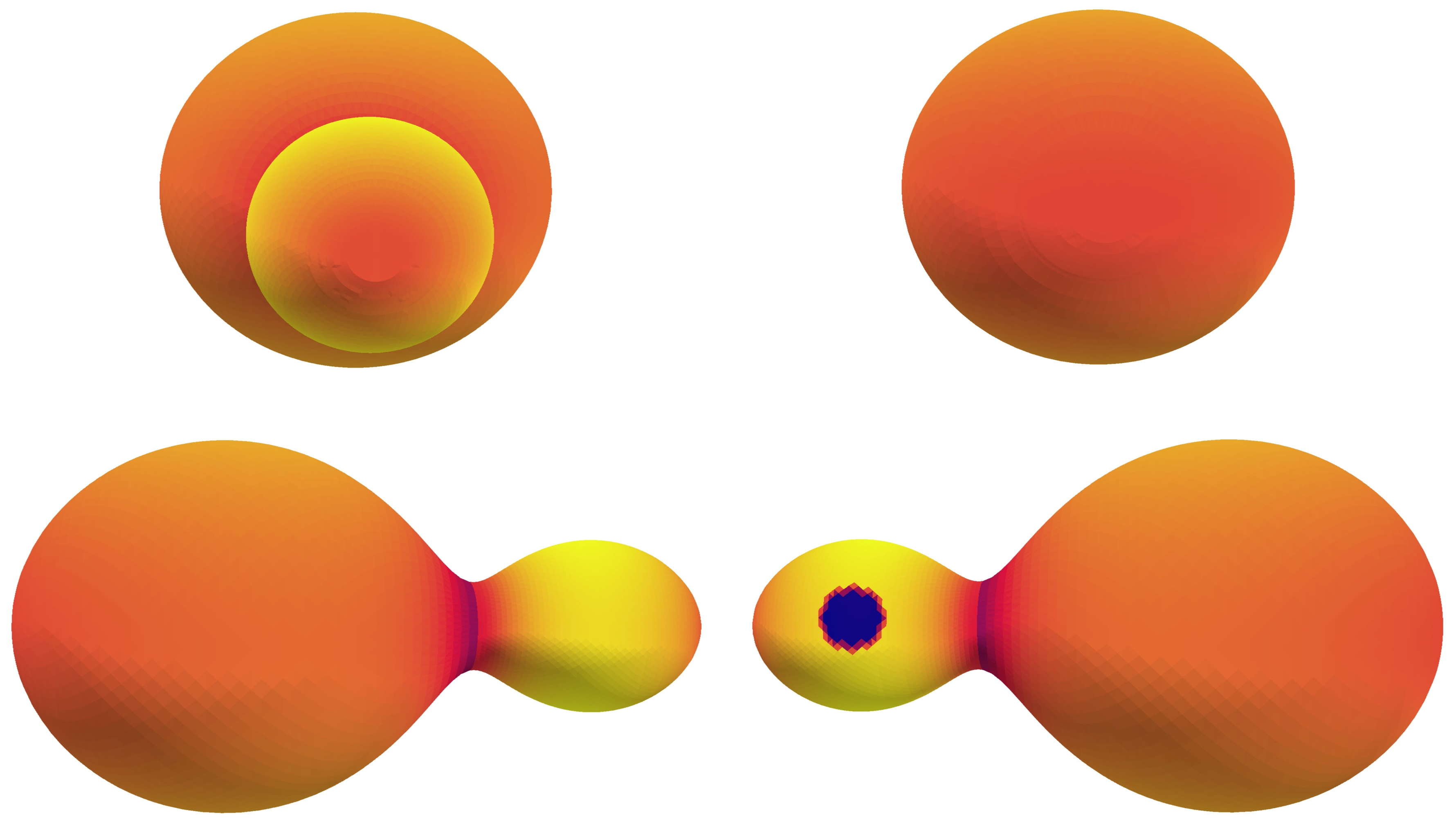}
\caption {Three-dimensional view of the EL Tuc based on its light curve solution.}
\label{Fig5}
\end{center}
\end{figure*}

The light curve analysis parameters' outcomes and uncertainties can be found in Table \ref{tab3}. Figure \ref{Fig3} shows the observed and theoretical light curves in different filters. The corner plot generated by MCMC is shown in Figure \ref{Fig4}. A cold starspot can be observed on the hotter component at phase 0.75, and the geometric structure of EL Tuc is illustrated in Figure \ref{Fig5}. In this system, with a large fillout factor of $f=0.537$, both stars overfill their inner Roche lobes, categorizing it as an overcontact binary. Additionally, the system's light curve analysis revealed no evidence of a third object light ($l_3$).

\begin{table*}
\caption{Photometric solution of EL Tuc.}
\centering
\begin{center}
\footnotesize
\begin{tabular}{c c c c c c c}
 \hline
 \hline
Parameter && Result && Parameter && Result\\
\hline
$T_{1}$ (K) && $5639_{\rm-20}^{+15}$ && $r_{1(mean)}$ && $0.558\pm0.003$\\
\\
$T_{2}$ (K) && $5825_{\rm-18}^{+16}$ && $r_{2(mean)}$ && $0.266\pm0.003$\\
\\
$q=M_2/M_1$ && $0.172_{\rm-0.001}^{+0.003}$ && Phase shift && $-0.012\pm0.001$\\
\\
$\Omega_1=\Omega_2$ && $2.104\pm0.054$ && Colatitude$_{spot}$(deg) && $99.630_{\rm-0.730}^{+0.640}$\\
\\
$i^{\circ}$ &&	$83.74_{\rm-0.42}^{+0.39}$ && Longitude$_{spot}$(deg) && $274.920_{\rm-0.560}^{+0.580}$\\
\\
$f$ && $0.537_{\rm-0.017}^{+0.016}$ && Radius$_{spot}$(deg) && $18.130_{\rm-0.320}^{+0.390}$\\
\\
$l_1/l_{tot}$ && $0.794_{\rm-0.001}^{+0.001}$ && $T_{spot}/T_{star}$ && $0.799_{\rm-0.003}^{+0.004}$\\
\\
$l_2/l_{tot}$ && $0.205\pm0.001$ && Component$_{spot}$ && Hotter\\
\hline
\hline
\end{tabular}
\end{center}
\label{tab2}
\end{table*}

%%%%%%%%%%%%%%%%%%%%%%%%%%%%%%%%%%%%%%%%%%%%%%%%%%
\vspace{1cm}
\section{Absolute Parameters}
There are various methods for estimating the absolute parameters of contact binary stars using photometric data. One of the methods involves utilizing the Gaia DR3 parallax (\citealt{kjurkchieva2019w}, \citealt{li2021photometric}, \citealt{poro2024estimatingGaia}). Another method is to use the $P-a$ empirical parameters relationship (\citealt{li2022extremely}, \citealt{2024PASP..136b4201P}). Both methods were employed in this investigation to estimate and compare the absolute parameters of EL Tuc. In the following, each of these methods is explained separately.

The study \cite{poro2024estimatingGaia} provides in-detail information for utilizing the Gaia DR3 parallax to estimate absolute parameters. The process of estimation necessitates the system's distance from Gaia DR3, the extinction coefficient $A_V$, $V_{max}(mag.)$, $l_{1,2}/l_{tot}$, $BC_{1,2}$, $T_{1,2}$, $r_{mean_{1,2}}$, and $P(day)$. Using these parameters, we can determine $M_{V(system)}$, $M_{V_{1,2}}$, $M_{bol_{1,2}}$, $R_{1,2}$,$L_{1,2}$, $a_{1,2}$, and $M_{1,2}$, respectively.
The value of semi-major axis $a(R_{\odot})$ is derived using $a_{1,2}=R_{1,2}/r_{mean1,2}$. Parameter $a(R_{\odot})$ is determined from the average values of $a_1(R_{\odot})$ and $a_2(R_{\odot})$. We utilized $V_{max}=14.57\pm0.21$ from our observations, the extinction coefficient $A_V=0.062\pm0.001$ calculation from the 3D dust map (\citealt{2019ApJ...887...93G}), and the system's distance from Gaia DR3 $d_{(pc)}=0.767\pm0.019$.
The results of the Gaia DR3 parallax method (Method 1) for estimating the absolute parameters of the EL Tuc system are given in Table \ref{tab3}.

We also estimated the absolute parameters of the EL Tuc using the $P-a$ relationship (Method 2). The Equation \ref{eq3} presented an empirical relationship between the orbital period and semi-major axis from the \cite{2024PASP..136b4201P} study,

\begin{equation}\label{eq3}
a=(0.372_{\rm-0.114}^{+0.113})+(5.914_{\rm -0.298}^{+0.272})\times P
\end{equation}

where $P$ is the system's orbital period. We considered average uncertainty in Equation \ref{eq3} for calculating $a(R_{\odot})$. Subsequently, we determined each star's mass and uncertainty by applying the well-known Kepler's third law equation and the mass ratio derived from the light curve solution. Then, we calculated the other absolute parameters ($R$, $L$, $M_{bol}$, $g$) for each component. Also, the orbital angular momentum ($J_0$) of the system was estimated using the \cite{2006MNRAS.373.1483E} study (Equation \ref{eq4}).

\begin{equation}\label{eq4}
J_0=\frac{q}{(1+q)^2} \sqrt[3] {\frac{G^2}{2\pi}M^5P}
\end{equation}

where $q$ is the mass ratio, $M$ is the total mass of the system, $P$ is the orbital period, and $G$ is the gravitational constant. The outcome of estimating the absolute parameters is presented in Table \ref{tab3}.

\begin{table*}
\caption {Estimation of the absolute parameters of the EL Tuc contact binary system.}
\centering
\begin{center}
\footnotesize
\begin{tabular}{c c c c c}
 \hline
 \hline
Parameter &\multicolumn{2}{c}{Method 1} & \multicolumn{2}{c}{Method 2}\\
 & Star1 & Star2 & Star1 & Star2\\
\hline
$M(M_\odot)$ & 1.630±0.419 & 0.294±0.076 & 1.334±0.394 & 0.230±0.073\\
$R(R_\odot)$ & 1.426±0.172 & 0.700±0.066 & 1.320±0.127 & 0.629±0.065\\
$L(L_\odot)$ & 1.793±0.409 & 0.519±0.092 & 1.589±0.347 & 0.411±0.095\\
$M_{bol}(mag.)$ & 4.106±0.245 & 5.453±0.192 & 4.237±0.215 & 5.705±0.226\\
$log(g)(cgs)$ & 4.342±0.215 & 4.216±0.192 & 4.321±0.033 & 4.202±0.034\\
$a(R_\odot)$ & \multicolumn{2}{c}{2.560±0.217} & \multicolumn{2}{c}{2.366±0.214}\\
$log(J_0)$ & \multicolumn{2}{c}{51.51±0.17} & \multicolumn{2}{c}{51.36±0.19}\\
\hline
\hline
\end{tabular}
\end{center}
\label{tab3}
\end{table*}

%%%%%%%%%%%%%%%%%%%%%%%%%%%%%%%%%%%%%%%%%%%%%%%%%%
\vspace{1cm}
\section {Discussion and Conclusion}
We employed ground-based photometric observations using $BVR_cI_c$ filters to investigate EL Tuc, a southern hemisphere eclipsing binary system. The summary, discussion, and conclusion of the obtained results in this investigation are as follows:
\\
\\
A) We extracted eight minima times from our photometric data. By supplementing these with the minima from the literature, we presented a new ephemeris for the EL Tuc system. Due to the limited data points on the O-C diagram, employing a linear least squares fit on the O-C diagram was appropriate.
\\
\\
B) We utilized the PHOEBE Python code for light curve analysis and employed MCMC to derive the final parameters' values and uncertainties. We used the initial effective temperature on the hotter component from the Gaia DR3 databases. The effective temperature difference between the two companion stars is 186 K, and the secondary star is the hotter component. Also, based on the stars' temperatures and the study of \cite{cox2015allen} and \cite{eker2018interrelated}, we have ascertained that the cooler star belongs to the G7 spectral type and the hotter component is classified as G3. It should be noted that the light curve analysis of EL Tuc required the addition of a cold starspot on the hotter component.
\\
\\
C) \cite{pribulla2003catalogue} found that Photometric ($q_{ph}$) and spectroscopic ($q_{sp}$) mass ratios for contact binaries could have nearly similar accuracy for the total eclipsing systems. This result implies that photometric observations can determine mass ratios accurately for total eclipsing binaries.

\begin{figure*}
\begin{center}
\includegraphics[width=0.75\textwidth]{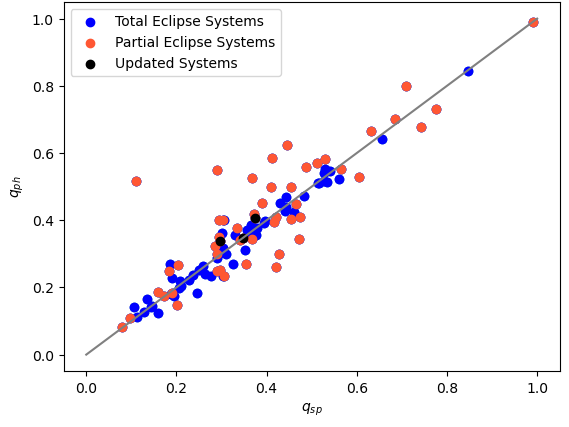}
\caption{The relation between the spectroscopic and photometric mass ratios. Based on the most recent studies, the black dots show the photometry mass ratios of the binary systems that have been updated.}
\label{Fig6}
\end{center}
\end{figure*}

As shown in Figure \ref{Fig6}, the $q_{ph}$ and $q_{sp}$ of 94 eclipsing binary systems presented by \cite{li2021photometric}. We added thirteen contact systems listed in Table \ref{tab4}, that have both $q_{ph}$ and $q_{sp}$. We have updated $q_{ph}$ for three systems listed in the \cite{li2021photometric} study, based on the latest photometric studies by black dots in Figure \ref{Fig6}. We determined if the contact systems are total or partial eclipsing binary by using the equating $i>arccos|\frac{r_{1(mean)}-r_{2(mean)}}{a}|$ (\citealt{sun2020physical}).

Figure \ref{Fig6} confirms that the results of mass ratio with photometric data can be reliable for total contact binaries, as the study of \cite{li2021photometric} had such a conclusion. According to EL Tuc is a total eclipse system, we employed the $q$-search method to find the initial mass ratio. The $q$-search plot depicts a deep and sharp minimum for our target system, and the MCMC process validates a $q=0.172$, confirming the EL Tuc system as an overcontact binary with a low mass ratio.

\begin{figure*}
\begin{center}
\includegraphics[width=0.8\textwidth]{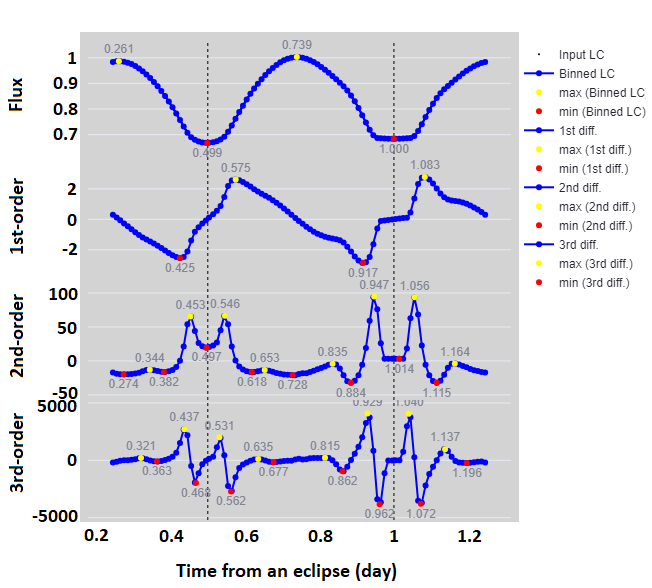}
\caption{The light curve of EL Tuc, and first to third derivatives (top to bottom panels), respectively. The units of the panels on the vertical axis from top to bottom are W m$^{-2}$, 10 W m$^{-2}$ day$^{-1}$, $10^2$ W m$^{-2}$ day$^{-2}$, and $10^4$ W m$^{-2}$ day$^{-3}$, respectively.}
\label{Fig7}
\end{center}
\end{figure*}

\begin{figure*}
\begin{center}
\includegraphics[width=0.86\textwidth]{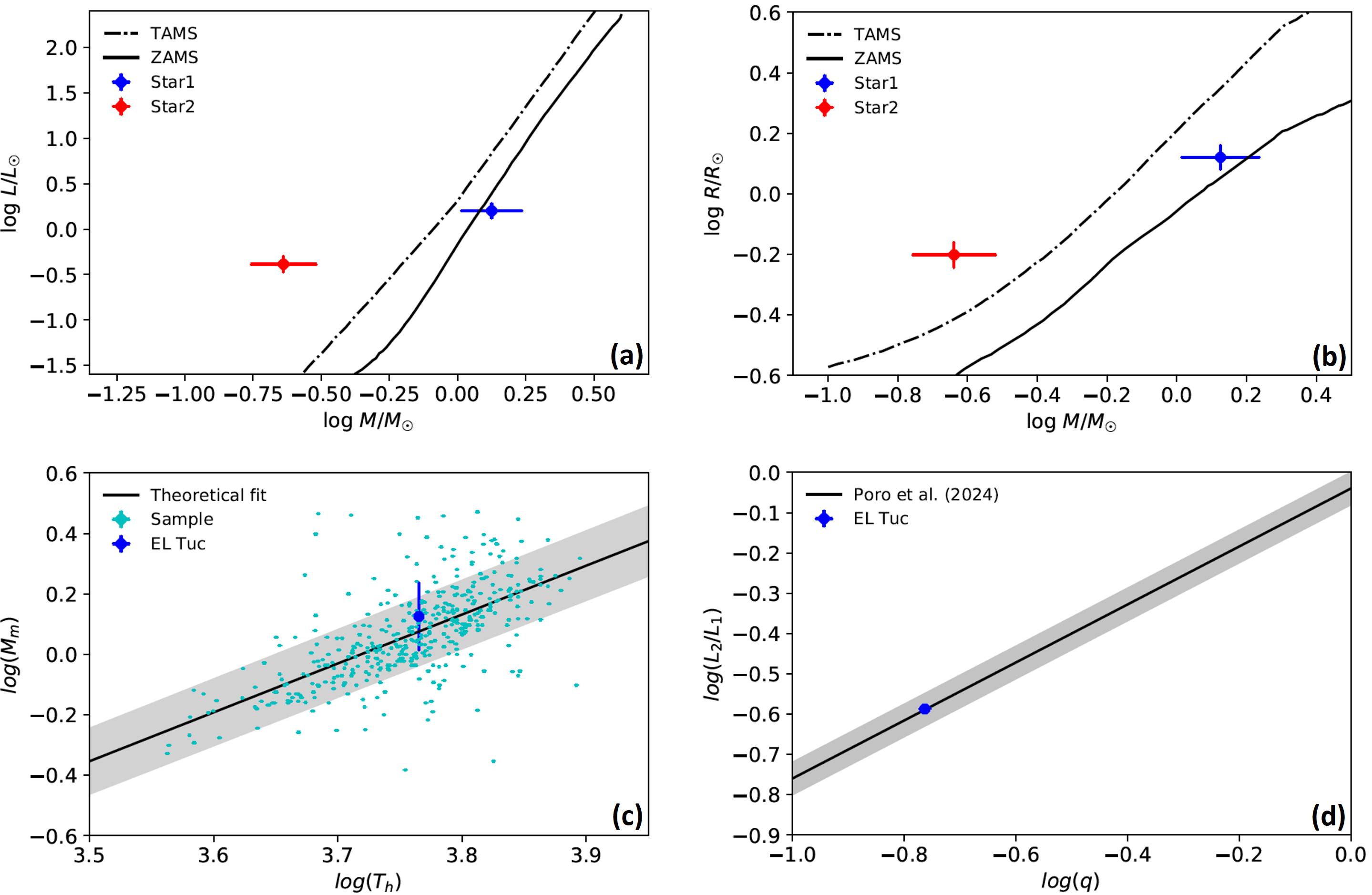}
\caption{The $\log{M} - \log{L}$, $\log{M} -\log{R}$, $\log{T_{h}} - \log{M_{m}}$, and $\log{q} - \log{L_{ratio}}$ diagrams.}
\label{Fig8}
\end{center}
\end{figure*}

However, we examined the mass ratio result from our light curve solution using another new and different method. Recently, \cite{2023ApJ...958...84K} proposed a novel method to use derivatives of the light curve to estimate the photometric mass ratio of overcontact binaries. Since iterative methods are typically employed to estimate the photometric mass ratio or take into consideration the relationships between the parameters, this method could be different from others. Based on \cite{2023ApJ...958...84K} study, this new method can be used for most overcontact systems. The discussed method is based on derivation at different orders of the photometric light curve. After deriving the light curve, at least two maxima should be seen in the light curve. Following the third-order derivation, a parameter called $W$ is obtained based on the measurement between the system's orbital period and the current minima and maxima. The analysis conducted by \cite{2023ApJ...958...84K} showed a strong relationship between $W$ and $q$. We used this method for EL Tuc, and the result was a mass ratio of $q=0.157\pm0.043$. Our light curve analysis and the \cite{2023ApJ...958...84K} study yielded a close mass ratio estimate with a discrepancy of 0.015. Therefore, considering the uncertainties, the results are in good agreement. Figure \ref{Fig7} shows the process of derivation of the EL Tuc light curve.
\\
\\
D) We estimated the absolute parameters of the system using two methods (Table \ref{tab3}). One of the methods was the use of the Gaia DR3 parallax, and the other was the empirical $P-a$ parameter relationship. It seems that the results of method 2 have better agreement with the theoretical relationships (Figure \ref{Fig8}); however, we have presented the results of both methods for consideration in future studies on this system.

We plotted the positions of the components in the Mass-Luminosity $(M-L)$ and Mass-Radius $(M-R)$ diagrams with the Zero-Age Main Sequence (ZAMS) and the Terminal-Age Main Sequence (TAMS) based on the estimated absolute parameters by Method 2 (Table \ref{tab3}, Figure \ref{Fig8}a,b). Figure \ref{Fig8}a,b illustrates that the cooler component is situated close to ZAMS, whereas the hotter star is located above TAMS.

Based on the \cite{poro2023two} study, the temperature-mass ($T_{h}-M_{m}$) relationship for contact binary systems with a linear fit (Equation \ref{eq5}) is shown in Figure \ref{Fig8}c, where $M_m$ represents the more massive component. The position of our target system on the $T_{h}-M_{m}$ diagram aligns well with the other sample stars from the \cite{latkovic2021statistics} study.

\begin{equation}\label{eq5}
\log M_{m} = (1.6185 \pm 0.0150) \times \log T_{h} + (-6.0186 \pm 0.0562)
\end{equation}

The systems' positions on the $q-L_{ratio}$ relationship from \cite{poro2023two} are also shown in Figure \ref{Fig8}d. The position of the EL Tuc system in this diagram is in good agreement with the theoretical fit.
\\
\\
E) According to the low mass ratio $q=0.172$, a high fillout factor $f=53.7\%$, and the orbital inclination of $i=83.74$, we can conclude that EL Tuc is a total overcontact binary system. Based on the results of the light curve analysis and the estimation of absolute parameters, EL Tuc is categorized as a W-subtype since the less massive component has a higher effective temperature.

\begin{table*}
\caption{The contact binaries with spectroscopic and photometric mass ratios.}
\centering
\begin{center}
\footnotesize
\begin{tabular}{c c c c c c c c c}
 \hline
 \hline
Name & {P(days)} &{$i^{\circ}$} &{$q_{sp}$} &{$q_{ph}$} & References\\
\hline
1RXS J034500.5+493710 & 0.37514 & 60 & 0.421 & 0.261 & \cite{kjurkchieva2019photometric} \cite{ding2023fundamental}\\
BD+42 765 & 0.35168 & 65 & 0.296 & 0.337 & \cite{kjurkchieva2019photometric} \cite{poro2023two}\\
V1191 Cyg & 0.31338 & 80 & 0.107 & 0.141 & \cite{ulacs2012marginally} \cite{ding2023fundamental}\\
V1073 Cyg & 0.78585 & 68 & 0.303 & 0.401 & \cite{tian2018multi} \cite{ding2023fundamental}\\
V402 Aur & 0.60349 & 63 & 0.201 & 0.148  & \cite{pych2004radial} \cite{ding2023fundamental}\\
DN Cam & 0.49830 & 73 & 0.421 & 0.411 & \cite{baran2004physical}
\cite{ding2023fundamental}\\
BO Ari & 0.31819 & 82 & 0.19& 0.227 & \cite{guerol2015absolute} \cite{ding2023fundamental}\\
DY Cet & 0.44079 & 82 & 0.356 & 0.370 & \cite{deb2011physical} \cite{ding2023fundamental}\\
AD Phe & 0.37992 & 76 & 0.367 & 0.344 & \cite{pi2017magnetic} \cite{ding2023fundamental}\\
FP Boo & 0.64048 & 69 & 0.096 & 0.109 & \cite{gazeas2006physical} \cite{ding2023fundamental}\\
EL Aqr & 0.48141 & 70 & 0.203 & 0.266 & \cite{deb2011physical} \cite{ding2023fundamental}\\
SX Crv & 0.31659 & 61 & 0.079 & 0.082 & \cite{zola2004physical} \cite{ding2023fundamental}\\
LS Del & 0.36384 & 48 & 0.375 & 0.407  & \cite{pych2004radial} \cite{poro2024bsn}\\
XZ Leo & 0.48773 & 78 & 0.348 & 0.346 & \cite{pych2004radial} \cite{luo2015photometric}\\
\hline
\hline
\end{tabular}
\end{center}
\label{tab4}
\end{table*}

%%%%%%%%%%%%%%%%%%%%%%%%%%%%%%%%%%%%%%%%%%%%%%%%%%
\vspace{1cm}
\section*{Data Availability}
Ground-based data will be made available on request.

%%%%%%%%%%%%%%%%%%%%%%%%%%%%%%%%%%%%%%%%%%%%%%%%%%
\vspace{1cm}
\section*{Acknowledgements}
This manuscript was prepared by the BSN project. We have made use of data from the European Space Agency (ESA) mission Gaia (\url{http://www.cosmos.esa.int/gaia}), processed by the Gaia Data Processing and Analysis Consortium (DPAC).

%%%%%%%%%%%%%%%%%%%%%%%%%%%%%%%%%%%%%%%%%%%%%%%%%%
\vspace{1cm}
\section*{ORCID iDs}
\noindent Elham Sarvari: 0009-0006-1033-5885\\
Eduardo Fernández Lajús: 0000-0002-9262-4456\\
Atila Poro: 0000-0002-0196-9732\\

%%%%%%%%%%%%%%%%%%%%%%%%%%%%%%%%%%%%%%%%%%%%%%%%%%
\vspace{1cm}
\bibliography{Ref}{}
\bibliographystyle{aasjournal}

\end{document}